\documentclass{article}[12pt]
\usepackage[margin=2cm]{geometry} 
\usepackage{graphicx}%
\usepackage{multirow}%
\usepackage{amsmath,amssymb,amsfonts,bbm}%
\usepackage{amsthm}%
\usepackage{mathrsfs}%
\usepackage{xcolor}%
\usepackage{booktabs}%
\usepackage{algorithm}%
\usepackage{algorithmicx}%
\usepackage{algpseudocode}%

\usepackage{listings}%
\newcommand{\Dcal}{\mathcal{D}}
\newcommand{\Ecal}{\mathcal{E}}
\newcommand{\Fcal}{\mathcal{F}}

\newcommand{\Mcal}{\mathcal{M}}
\newcommand{\Ncal}{\mathcal{N}}

\newcommand{\Scal}{\mathcal{S}}

\newcommand{\Xcal}{\mathcal{X}}

\newcommand{\R}{\mathbb{R}}
\newcommand{\E}{\mathbb{E}}
\newcommand{\Vbb}{\mathbb{V}}
\newcommand{\Pbb}{\mathbb{P}}

\newcommand{\one}{\mathbbm{1}}
\newcommand{\angleavg}[1]{ \langle {#1} \rangle }

\newtheorem{theorem}{Theorem}
\newtheorem{proposition}[theorem]{Proposition}%
\newtheorem{remark}[theorem]{Remark}%
\newtheorem{corollary}[theorem]{Corollary}

\begin{document}
\title{Model free collision aggregation for the computation of escape distributions}

\author{Laetitia Laguzet\thanks{CEA-DAM-DIF, F-91297 Arpajon, France, \text{Latitia.Laguzet@cea.fr},  ORCID  0009-0000-9655-0880}
\and Gabriel Turinici\thanks{CEREMADE, Université Paris Dauphine - PSL,  Paris 75116, Paris, France; \text{Gabriel.Turinici@dauphine.fr}, \text{https://turinici.com}, ORCID 0000-0003-2713-006X}}
\date{Sept 2024}
\maketitle

\begin{abstract} Motivated by a heat radiative transport equation, we consider a particle undergoing collisions in a space-time domain and propose a method to sample its escape time, space and direction from the domain. 
The first step of the procedure is an estimation of how many elementary collisions is safe to take before chances of exiting the domain are too high; then these collisions are aggregated into a single movement. The method does not use any model nor any particular regime of parameters.
We give theoretical results both under the normal approximation and without it and test the method on some benchmarks from the literature. The results confirm the theoretical predictions and show that the proposal is an efficient method  to sample the escape distribution of the particle.


{\bf keywords : heat radiative transfer equation; Monte Carlo simulation; collision transport equations}

Subject classification : 65C05, 80M31, 35Q49%

\end{abstract}

\section{Introduction}

Particle simulations offering insights into complex chemical systems at the molecular level and can  help elucidate reaction mechanisms, predict thermodynamic properties, and explore molecular assembly; such methods have been successfully applied in various fields such as catalysis \cite{doi:10.1021/jp962095q},
 atmospheric modelling \cite{hersch_radiative_transport_jpca,hersch_radiative_transport20}, radiation transport \cite{fleck1984random}, etc. 
We will focus on the integro-differential transport equation:
\begin{equation}
	\frac{1}{c} \partial_t u(t,x,\omega) + \omega \cdot \nabla u(t,x,\omega) + (\sigma_a(t,x) + \sigma_s(t,x)) u(t,x,\omega) = \sigma_s(t,x) \angleavg{u}(t,x),
	\label{eq:transport}
\end{equation}
with time variable $t \in [0,T]$, position variable $x\in \mathcal \Xcal \subset \R^d$ (here $d\ge 1$ is the spatial dimension), the angle of propagation $\omega \in \mathcal{S}^{d}$ (unit sphere in $\R^d$) and $\angleavg{u}(t,x) = \frac{\int_{\mathcal{S}^d} u(t,x,\omega') d \omega'}{\int_{\mathcal{S}^d} 1 \cdot d \omega'}$
the angular average of $u$ on $\mathcal{S}^d$. 
The  model represents heat radiative transfer equations that can be used for both photons and neutrons (we use the former in the numerical results).

The model also comes with the
constant speed $c$~; for instance for photons this will be the speed of light. The absorption opacity  $\sigma_a$ and the scattering opacity $\sigma_s$ are (known) functions of $x$ and $t$ that describe the collision dynamics of the particles and more precisely the time to next collision, see section~\ref{sec:simple_model} for details. 

The spatial domain $\Xcal$ is usually a mesh simulation cell and various approximations are invoked to compute relevant quantities and manage the transition of particles from one mesh cell to another. We will not discuss this but refer to \cite{FLECK1971313,brooks1989symbolic} and related literature. The main focus of this paper will be on how to compute the evolution of one particle from the initial time $t_0=0$ and initial position $x_0 \in \Xcal$ to the moment when it exits the domain $\Xcal \times [0,T]$, i.e. either reaches the spatial boundary $\partial \Xcal$ or consumes all available time $T$.

We are concerned here with `Monte Carlo' approaches that regard~\eqref{eq:transport} as the time-evolving probability density of a stochastic process, see~\cite{lapeyre1998methodes}.  When the parameter $\sigma_s$ is large, many collisions occur before final time $T$ because the average time to next collision is $\frac{1}{\sigma_s}$. This is the so-called `diffusion regime'~\cite{larsen1980diffusion} and approximation methods exist to exploit this remark, in particular the  
Random Walk (RW) methods \cite{fleck1984random, giorla1987random}. 
On the contrary, when $\sigma_s$ is small, the particle will not undergo many collisions before exiting the mesh cell and the ballistic regime is important. In between, there are situations where the diffusion limit is not valid but the number of collisions is still important and requires extensive numerical simulations. Like in the diffusion limit, one would like to somehow accelerate this computation by replacing a large sum of independent collisions with some aggregate step, without resorting to diffusive approximations. So we focus in this paper on a {\bf model free} method to aggregate many collisions into a single displacement without affecting the escape distribution of the particle. The simulation of the particle's trajectory stops when either time ends (at $t=T$) or the particle reaches the spatial boundary $\partial \Xcal$. Note that a single aggregated step will probably not suffice to end the simulation for the particle so several such movements will probably be used.

As a technical circumstance, we will invoke the discrete ordinate method, denoted  $S_N$, which consists in discretizing 
the angular direction variable i.e., replacing $\Scal^d$ by a set of discrete directions $\Dcal$.

The outline of the paper is the following: we describe the collision dynamics of one particle in section \ref{sec:simple_model}. Then in section~\ref{sec:idea} we describe the general principle of the proposed method; in sections~\ref{sec:estimatestn} and \ref{sec:estimatesx} we give several  theoretical results. The procedure is then tested in section~\ref{sec:numerical}.

\section{Procedure and associated theoretical insights}

\subsection{Description of the collision dynamics}
\label{sec:simple_model}

We present briefly the dynamical setting. For graphical convenience this problem is presented in 1D but it transcribes without difficulty to the multi-dimensional case.

For further simplification we will restrict to a situation where the direction of the particle is either $+1$ or $-1$; this is the so-called $S_2$
discrete-ordinates method
(see \cite[section 16.3 page 502]{modest2021radiative},  which relates to the Schuster-Schwarzschild equations \cite[section 14.3 p 456]{modest2021radiative}).
We will denote $\Dcal_2= \{-1,1\}$ and in general 
\begin{equation}
\Dcal_N = \left\{cos\left(\frac{2 \pi j}{N} \right), j=0,...,N-1\right\}.
\label{eq:def_sn_directions}
\end{equation}
Note that, for dimensions $d$ higher than $1$, the direction of the particle is an element of the unit sphere $\Scal^d$. 
We consider a $1D$ particle  in the spatial domain $\Xcal := [0,L]$ starting from 
$x(0)=x_0 \in \Xcal$ at $t_0=0$. We are also given a maximum time $T$.
The particles evolve as follows : the collision counter is set to $s=1$
and a direction $a_s\in \Dcal$ is chosen at random uniformly from $\Dcal$. 
Here $\Dcal=\Dcal_N$ and unless stated otherwise we set $N=2$ but our most general results in section
\ref{sec:theory_no_gauss_generalN}
 consider arbitrary values of $N$.

 A  time $\tau_s$ is sampled from an exponential law of mean $1/\sigma$ (this will be denoted $\tau_s \sim Exp(\sigma)$). We define $t_s=t_{s-1}+ \tau_s$.
The particle moves on a straight line in the direction $a_s$ during the time $\tau_s$ at constant speed $v$. Thus for any $t \in [t_{s-1},t_s]$ : $x(t) = x(t_{s-1})+ v(t-t_{s-1}) a_s$.

Next, the collision counter $s$ is incremented $s \to s+1$ and the process repeats
until either $t_{s^\star}\ge T$ or $x_{t_{s^\star}} \notin \Xcal$. The precise space-time coordinates when the particle touches the first time the boundary of the domain $\Xcal \times [0,T]$ are computed, i.e., for our simple 1D case the smallest $t^\star \in [t_{{s^\star}-1},t_{s^\star}]$ such that $x(t^\star)=0$ or $x(t^\star)=L$ or $t^\star=T$.

The quantity of interest is the distribution of the escape space-time, more precisely the joint distribution of the escape position  $x^\star:=x(t^\star)$, escape direction $a^\star \in \Dcal$ and escape time $t^\star$ when the boundary is reached.
This triplet $(x^\star,t^\star,a^\star)$ is a random variable whose distribution $\Ecal(\sigma,\Xcal,T,x_0)$ depends only on $\sigma$, $\Xcal$,  $T$ and $x_0$.

\begin{figure}[htb!]
\begin{center}
\includegraphics[width=.49\textwidth]{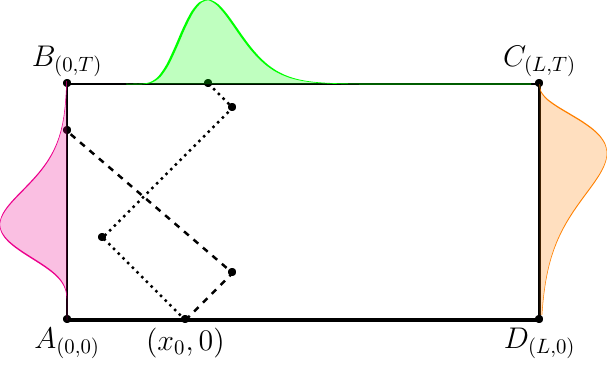}	
\caption{An illustration of the time dynamics and escape space-time distribution for  a particle starting at $x_0$, $\Dcal=\Dcal_2$. 
The abscissas represent the position of the particle and the ordinates the time.
For each remarkable point we give its $x,t$ coordinates, for instance $A$ has $x=0$ and $t=0$ and represents the left extremity of the segment at $t=0$ while $C$ is the left extremity at the final time. 
The time to next collision is an exponential random variable of average $1/\sigma$ i.e., distributed  $Exp(\sigma)$. The joint escape distribution $\Ecal(\sigma,\Xcal,T,x_0)$ collects the location $x^\star$, time $t^\star$ and direction $a^\star$ when the boundary $x=0$, $x=L$ or $t=T$ is reached for the first time (i.e., particle hits $AB$, $BC$ or $CD$).	
The simulation is stopped when this happens. The colored histograms are artist views of each conditional escape distributions: the magenta/orange is the histogram of the escape {\bf time $t^\star$} given that particle escaped through the left/right while the green is the histogram of escape {\bf position $x^\star$} for particles that did not escape before time $T$ or, equivalently, the escape is due to time $T$ being totally consumed.}  \label{fig:xminxmaxtmax}
	\end{center}
\end{figure}

An illustration is given in figure~\ref{fig:xminxmaxtmax} for $\Dcal = \Dcal_2$.
Note that the escape space-times values are elements of $\R^{2 dim(\Xcal)+1}= \R^3$
thus the support of the distribution 
$\Ecal(\sigma,\Xcal,T,x_0)$ is included in $R^3$; but there are many restrictions on this support, for instance $t\in [0,T]$ and so on, leaving the support to be included in (see figure \ref{fig:xminxmaxtmax} for notations):
\begin{equation}
	\Big( AB \times \{-1\} \Big)  \bigcup \Big(BC \times \{-1,1\} \Big) 
	\bigcup \Big( CD \times \{1\} \Big).
\end{equation}
For instance, the explanation of the first element $ AB \times \{-1\} $ is that if the particle exists through its left border the exit direction will point to the left. Or in $\Dcal$ there is only one direction that points to the left which is $-1$;  same for $CD \times \{1\}$.

The colored areas in figure~\ref{fig:xminxmaxtmax} are a general illustration, not corresponding to any specific parameters, of the following three conditional distributions: 

- the left area (magenta in color figure) is the distribution of the escape {\bf time} $t^\star$ 
at which $x(t^\star)=0$, 
conditioned by the fact that the particle touched $x=0$ before $T$ and before touching $x=L$ ;

- the top area (green in color figure) is the distribution of the escape {\bf position} $x(t^\star)$ conditioned by escaping because time $T$ was reached before reaching $x=0$ or $x=L$, i.e.  $t^\star=T$,

- the right area (orange in color figure) is the equivalent to the magenta area when border is first reached for $x=L$.

\subsection{Idea and first estimations} \label{sec:idea}

The dynamics described in section \ref{sec:simple_model} is used to sample from
the distribution $\Ecal$. In particular the trajectory 
 before the exit time i.e., $x(t)$ for $t<t^\star$, 
is not useful and not used.
We can thus imagine a way to accelerate the computation by "skipping" these intermediary steps. For instance, when the diffusion parameter $\sigma$ is very large many collisions will occur before particle exits the space-time domain and in this case a random walk approximation could be valid. We do not want to use this kind of approximation here but remain as close as possible to the collisional dynamics.

Recall that the collision time $t_n$ is the sum of i.i.d $Exp(\sigma)$ random variables $\tau_j$: 
 $t_n = \sum_{j=1}^n \tau_j$; the law of $t_n$ is a Gamma distribution
 of parameters $n$ and $\sigma$.

The position $X_n= x(t_n)$ is such that   
$X_n = x_0+\sum_{j=1}^n v a_j \tau_j$, its law is a mixture of sums of two Gamma distributed random variables (one for each value in $\Dcal$); we will make this precise latter. In any case we will show that for any $n$ one can sample {\bf directly and exactly} from the joint law $(X_n,t_n)$. In this way we can advance the time by $t_n$ units and replace $n$ individual collisions by only one sample from this joint law.

But the question that arises is the following: what is the value of $n$ such that, with high certainty, we can make 
$n$ steps {\bf without exiting the space-time domain $\Xcal\times [0,T]$} ? 
Given some tolerance $\epsilon >0$ 
the overall goal of the estimations in following sections is to find how large $n$ can be while still ensuring that:
\begin{equation}
\Pbb[ \{ \exists t \le t_n : x(t) \notin \Xcal \} \cup \{t_n \ge T \}] \le \epsilon. 
\end{equation}

\begin{remark}
The event $\{ \exists t \le t_n : x(t) \notin \Xcal \} \cup \{t_n \ge T \}$ whose probability is evaluated above concerns trajectories where either $t_n$ is larger than $T$ or trajectory already exited the domain $\Xcal$ before $t_n$. The complementary of this event are trajectories $(x(t),t)$ that remained in  
$\mathcal{X}\times [0,T]$
for all $t \le t_n$.  For trajectories in the complementary 
of $\{ \exists t \le t_n : x(t) \notin \Xcal \} \cup \{t_n \ge T \}$
 we do not need to sample $x(t_1)$, ... $x(t_{n-1})$ because, with high probability, no exit occurred in $[0,t_n[$.
If $\epsilon$ is small then with probability at least $1-\epsilon$ one can thus only sample $t_n,x(t_n)$ (and the direction at time $t_n$). This is the goal of this estimation, to know how many intermediary steps $x(t_1)$, ... $x(t_{n-1})$ can be safely omitted. Note that $x(t_n)$ is still sampled exactly; the acceleration provided by our procedure comes from the fact that we do not need to sample $x(t_1)$, ..., $x(t_{n-1})$.
Then the procedure computes, starting from $x(t_n)$ another estimation for how many steps can be aggregated from this new position with very low risk of exiting the domain
$\Xcal\times [0,T-t_n]$ and so on.

Note that when $\epsilon$ is small this does not imply $n$ is also small, cf. for instance estimation 
\eqref{eq:normal_approx_tn}  below where $\epsilon=10^{-9}$ and $n$ can be as large as 
$ T\sigma +18 - 6\sqrt{T\sigma+9}$ (which is large provided $T\sigma$ is large, see table~\ref{table:valuestsigmaetc} for examples).    
\end{remark}

\subsection{Estimates for $t_n$} \label{sec:estimatestn}

\subsubsection{Normal approximation for $t_n$}
The first approach is to use a normal (Gaussian) approximation. 
This idea is close to the {\it Random Walk} regime (cf. Introduction). 
For instance we know that $t_n$ has mean $n/\sigma$ and variance $n/\sigma^2$. For very large values of $n$, $t_n$  will behave as a normal variable with same mean and variance, i.e. $ \frac{\sigma}{\sqrt{n}} (t_n -n/\sigma)$ is "close" to a standard normal.  
So, we will write
$\Pbb[ t_n \ge T] \le \epsilon$ is the same as 
$\Pbb[ \frac{\sigma}{\sqrt{n}} (t_n -n/\sigma) \ge  \frac{\sigma}{\sqrt{n}} (T -n/\sigma)] \le \epsilon$ 
which, if the normal approximation holds, will be true when 
$\frac{\sigma}{\sqrt{n}} (T -n/\sigma)$ is larger than the $1-\epsilon$ quantile of the normal distribution. To simplify things we take as small error
$\epsilon = 10^{-9}$
corresponding to 
slightly more than 
$6$ standard deviations (that we conservatively set to $6$). 
In this case, with high probability, $t_n$ will not be larger than $T$
if $\frac{\sigma}{\sqrt{n}} (T -n/\sigma) >6$
or, equivalently,  $n + 6 \sqrt{n} \le \sigma T$. So the aggregation rule
becomes :
\begin{equation}
	n \le T\sigma +18 - 6\sqrt{T\sigma+9}.
	\label{eq:normal_approx_tn}
\end{equation}
Note that here $T\sigma$ is the mean number of collisions required  to reach the final time $T$ (each collision "consumes" in average $1/\sigma$ time units).

\subsubsection{Exact tail approximation for $t_n$}

We can also give a more precise, but slightly less convenient, estimation for $n$ coming from the tail estimates for the gamma distribution. 
\begin{proposition}With previous notations, choosing $n \le T \sigma/2$:
	\begin{equation}
		\Pbb[ t_n \ge T] \le \frac{z^{z/2} e^{-z}}{\Gamma(1+z/2)}, \text{ where } 
		z=T \sigma.
	\end{equation}	
	In particular:
	\begin{equation}
		\Pbb[ t_n \ge T] \le 10^{-9}, \text{ if }
		n \le T\sigma/2, n\ge 50. \label{eq:tail_estimates_tn}
	\end{equation}

\end{proposition}
\begin{proof}
	We are interested in 
	$\Pbb[ t_n \ge T] = \Pbb[ \sigma t_n \ge T \sigma] $. Note that  
	$\sigma t_n$ is a gamma random variable with parameters shape=$n$ and scale=$1$ and density $f(y)=\frac{1}{(n-1)!} y^{n-1} e^{-y}$. We need to give a bound for 
	$\Pbb[ \sigma t_n \ge T \sigma] = \int_{T \sigma}^\infty \frac{1}{(n-1)! }y^{n-1} e^{-y} dy$. By integration by parts for general $z>0$:
	\begin{equation}
		\int_z^\infty \frac{y^{n-1}}{(n-1)! } e^{-y} dy = \frac{y^n}{n!}e^{-y}\Big|_{y=z}^\infty + 
		\int_z^\infty \frac{y^n}{n!} e^{-y} dy \ge 
		\frac{z}{n}
		\int_z^\infty \frac{y^{n-1}}{(n-1)! } e^{-y} dy - \frac{z^n}{n!}e^{-z}.
	\end{equation}
We used the inequality $\int_z^\infty \frac{y^n}{n!} e^{-y} dy \ge  \frac{z}{n}\int_z^\infty \frac{y^{n-1}}{(n-1)! } e^{-y} dy$ that is true because $y\ge z$ on the domain of integration.
The term $\int_z^\infty \frac{y^{n-1}}{(n-1)! } e^{-y} dy $ appears now in both sides and  thus one can write
$\left( \frac{z}{n} -1 \right) \int_z^\infty \frac{y^{n-1}}{(n-1)! } e^{-y} dy \le \frac{z^n}{n!}e^{-z}$; 
dividing by $\left( \frac{z}{n} -1 \right)$
 it follows that, for $z > n$:
	\begin{equation}
		\int_z^\infty \frac{y^{n-1}}{(n-1)! } e^{-y} dy \le	
		\frac{n z^n e^{-z}}{n! (z-n)} = \frac{n z^n e^{-z}}{\Gamma(n+1) (z-n)}.
	\end{equation}
	Let us take now $n=z/2$; the right hand side equals
	$\frac{z^{z/2} e^{-z}}{\Gamma(1+z/2)}$. The function 
	$z \mapsto \ln \left(\frac{z^{z/2} e^{-z}}{\Gamma(1+z/2)}\right) = 
	\frac{z}{2} \ln(z)-z- \ln(\Gamma(1+z/2))$ has derivative 
	$\frac{\ln(z)-1 - \psi(1+z/2)}{2}= 
 \frac{\ln(z/2)- \psi(1+z/2) + \ln(2/e)}{2} \le 0$ 
	we have used the notation for the digamma function $\psi(z) = \frac{\Gamma'(z)}{\Gamma(z)}$, 
	inequality $\ln(u) \le \psi(u+1)$ for all $u> 0$ and $\ln(2/e)<0$.
 This shows that the error term 
	$\frac{z^{z/2} e^{-z}}{\Gamma(1+z/2)}$ is decreasing for $z\ge 2$. In particular its value at $z=100$ is $2.44\times 10^{-10} < 10^{-9}$.
	Conclusion  \eqref{eq:tail_estimates_tn} follows because 
	$n\mapsto \Pbb[ t_n \ge T]$ is obviously increasing with $n$ and the value at $n = T\sigma/2$ is less than $10^{-9}$.
\end{proof}

\subsubsection{Summary for $t_n$}

To summarize, in order to satisfy $\Pbb[ t_n \ge T] < 10^{-9}$ we have two possible choices
\begin{itemize}
	\item the rigorous, conservative estimate from  \eqref{eq:tail_estimates_tn} with choice $n\le T \sigma/2$ as soon as $n\ge 50$
	\item the normal approximation \eqref{eq:normal_approx_tn} resulting in the bound:
	$	n \le T\sigma +18 - 6\sqrt{T\sigma+9}$.  In practice we require 
	$n$ to be larger than some $n_\Ncal$ that we set to $n_\Ncal:=300$
	 in order to ensure that the normal approximation is in the asymptotic regime.
\end{itemize} 
The normal approximation provides larger (thus less restrictive) values for $n$ but its quality is not precisely quantified. On the other hand, the conservative estimate
\eqref{eq:tail_estimates_tn} has a known error bound and works even if only one hundred average collisions are left in the time interval. 
If no other parameters enter into the decision, the number of such "aggregated collisions" required to reach final time $T$ is  $O(1)$
for the normal approximation and $O(\log(T \sigma))$ for the conservative estimate (each step halves the time "left"). Both give very good results for the 
numerical regimes we are interested in.
\subsection{Estimates for the spatial boundary} \label{sec:estimatesx}

We now inquire about estimating the number of collisions that can be aggregated without reaching the spatial boundary.
Define $D$ to be  the distance from $x_0$ to the boundary of $\Xcal$ i.e., 
$D= \min\{ x_0, L-x_0\}$. In general, if the spatial domain is $\Xcal$ we set 
$D = dist(x_0, \partial \Xcal)$.
Note that $\E[X_n-x_0]= 0$, $\Vbb[X_n-x_0]= n v^2 / \sigma^2$. But, this still does not tell us if some other $X_k$ for $k\le n$ did not already exited through the spatial domain boundaries $\{0\}$ or $\{L\}$. We can write:
\begin{equation}
	\Pbb[ \{ \exists t \le t_n : x(t) \notin \Xcal \}] \le 
	\Pbb[ \{ \max_{s\le n} |X_s-x_0| \ge D\} ]=
	\Pbb[ \{ \max_{s\le n} |\sum_{j=1}^s a_j \tau_j | \ge D/v\} ].
	\label{eq:proba_exit_before_tn}
\end{equation}

\subsection{Spatial boundary treatment, no Gaussian  approximation, general $N$}
\label{sec:theory_no_gauss_generalN}

 We will consider a general case when $\Dcal$ is not necessary $\{-1,1\}$.
  We will assume:
\begin{eqnarray}
& \ &  \text{Hypothesis on directions (HD)}:  
\nonumber \\ & \ & 
\forall j \ge 0: 
	\E[a_j]= 0, \Vbb[a_j] = \varsigma^2, \text{ the law of }  a_j \text{ is symmetric}.
\end{eqnarray}
In practice this can restrict for instance the number of directions to be even for the $S_N$ model (to ensure symmetry). Note that for $\Dcal=\Dcal_N$: $\varsigma= \sqrt{N/2}$. We will need the following result.
\begin{proposition} \label{prop:estimations}
Consider $Y_n= \sum_{j=1}^n Z_j$ with $Z_j$ symmetric i.i.d. variables such that 
for some $\lambda>0$: $\E[e^{\lambda Z_1}] < \infty$. Then for any $C\ge 0$
\begin{equation}
	\Pbb[ \max_{j \le n} |Y_j| \ge C] \le
	 4 \Pbb[ Y_n \ge C] = 
	 2\Pbb[ |Y_n| \ge C]  \label{eq:reflection_ineq}
\end{equation} 
\begin{equation}
\Pbb[ \max_{j \le n} |Y_j| \ge C] 
\le 2 e^{- \lambda C} \left( \E[ e^{\lambda Z_1}]\right)^n \label{eq:inequality_doob_exp}
\end{equation}
In particular $	\Pbb[ \max_{j \le n} |Y_j| \ge C] \le \epsilon$ as soon as:
\begin{equation}
	n \le \frac{\lambda C + \ln(\epsilon)- \ln(2)}{\ln( \E[ e^{\lambda Z_1}])}.
\label{eq:bound_n_exp}
\end{equation}

\end{proposition}
\begin{remark}
	The estimation \eqref{eq:reflection_ineq} reminds of the
	Kolmogorov's inequality that  would read:
\begin{equation}
	\Pbb[ \max_{j \le n} |Y_j| \ge C] \le  \frac{1}{C^2}\Vbb[Y_n] = \frac{n}{C^2}\Vbb[Z_1].
	\label{eq:reflection_ineq2}
\end{equation} 
Such an inequality is not useful because if would constraint $n$ to not be larger than $ \epsilon C^2/ \Vbb[Z_1]$ which is very disappointing when $\epsilon \to 0$. On the other hand, Doob's inequality will be invoked in the proof of the upper bound \eqref{eq:inequality_doob_exp} and  the estimation
\eqref{eq:bound_n_exp} where $\epsilon$ only appears through its logarithm.
\label{rem:markov}
\end{remark}

\begin{proof} {\bf Proof of inequality \eqref{eq:reflection_ineq}:} 
The general idea is that $Y_n$ can be thought close, by Donsker's reflection principle, to a Brownian motion; for a Brownian motion the reflection principle related the maximum deviation before time $t_n$ with the value at time $t_n$, i.e., the estimation  \eqref{eq:reflection_ineq} is true with equality. So we follow the proof of the Brownian motion reflection principle. Note first that 
\begin{eqnarray}
& \ & 	\Pbb[ \max_{j \le n} |Y_j| \ge C] = 	
\Pbb[ \{\max_{j \le n} Y_j \ge C\} \cup \{ \min_{j \le n} Y_j \le -C \} ] 
\le  	
\Pbb[ \{\max_{j \le n} Y_j \ge C\}] + \Pbb[ \{ \min_{j \le n} Y_j \le -C \} ] 
\nonumber \\ & \ & 
= \Pbb[ \{\max_{j \le n} Y_j \ge C\}] + \Pbb[ \{ C \le -\min_{j \le n} Y_j \} ]
= \Pbb[ \{\max_{j \le n} Y_j \ge C\}] + \Pbb[ \{ C \le \max_{j \le n} -Y_j \} ]
\nonumber \\ & \ & 
= 2\Pbb[ \{\max_{j \le n} Y_j \ge C\}],
\end{eqnarray}
where we used the symmetry of $Y_n$. By symmetry we also obtain 
\begin{equation}
\Pbb[ |Y_n| \ge C] = 2 \Pbb[ Y_n \ge C], \label{eq:yn_wo_abs}
\end{equation}
so it is enough to show that 
\begin{eqnarray}
	\Pbb[\max_{j \le n} Y_j \ge C] \le  2\Pbb[ Y_n \ge C]. \label{eq:relation_max_noabs}
\end{eqnarray}
Denote $u$ the stopping time to be the first $s$ such that $Y_s\ge C$ or $n$ if no such $s$ exists; define $\Fcal_u$ to be its associated sigma-algebra. Then: 
\begin{eqnarray}
& \ & \!\!\!\!\!\!\!\! \Pbb[\max_{j \le n} Y_j  \ge C] =\Pbb[ \max_{j \le n} Y_j  \ge C, Y_n \ge C ]+   
\Pbb[ \max_{j \le n} Y_j  \ge C, Y_n < C ] 
\nonumber \\ & \ &
\le  \Pbb[Y_n \ge C ]+\Pbb[ \max_{j \le n} Y_j \ge C, Y_n-Y_u < 0 ]
\nonumber \\ & \ &
= \Pbb[Y_n \ge C ]+\E[ \one_{\max_{j \le n} Y_j \ge C, Y_n-Y_u < 0} ]
\nonumber \\ & \ &
= \Pbb[Y_n \ge C ]+\E[ \E[\one_{\max_{j \le n} Y_j \ge C, Y_n-Y_u < 0} |\Fcal_u ]]
\nonumber \\ & \ &
= \Pbb[Y_n \ge C ]+\E[ \one_{\max_{j \le n} Y_j \ge C} \E[\one_{Y_n-Y_u < 0}]]
\nonumber \\ & \ &
= \Pbb[Y_n \ge C ]+\Pbb[ \max_{j \le n} Y_j \ge C]\cdot  \Pbb[Y_n-Y_u < 0]
\nonumber \\ & \ &
\le \Pbb[Y_n \ge C ]+\frac{1}{2}\Pbb[ \max_{j \le n} Y_j \ge C], 
\end{eqnarray}
where we used the fact that $Y_n-Y_u$ is symmetric thus 
$ \Pbb[Y_n-Y_u < 0] \le 1/2$. The relation \eqref{eq:relation_max_noabs} follows by subtracting $\frac{1}{2}\Pbb[ \max_{j \le n} Y_j \ge C]$ from the first and last terms of the formula above.

\noindent {\bf Proof of inequality \eqref{eq:inequality_doob_exp}:}
We already proved
$$\Pbb[ \max_{j \le n} |Y_j| \ge C] \le 2\Pbb[ \{\max_{j \le n} Y_j \ge C\}] = 
 2\Pbb[ \{\max_{j \le n} e^{\lambda Y_j} \ge e^{\lambda C}\}].$$ Since for $\lambda>0$, the function $x \mapsto exp(\lambda x)$ is convex, $e^{\lambda Y_n}$ is a sub-martingale. By Doob's inequality,
\begin{equation}
	\Pbb[ \{\max_{j \le n} e^{\lambda Y_j} \ge e^{\lambda C}\}]
\le e^{- \lambda C} \E[ e^{\lambda Y_n}] =  e^{- \lambda C} \prod_{j=1}^n \E[ e^{\lambda Z_j}]=  e^{- \lambda C} 
(\E[ e^{\lambda Z_1}])^n.
\end{equation}
\noindent {\bf Proof of inequality \eqref{eq:bound_n_exp}:}
it results from~\eqref{eq:inequality_doob_exp} by taking the logarithm
in $ 2 e^{- \lambda C} \left( \E[ e^{\lambda Z_1}]\right)^n \le \epsilon$.
\end{proof}

\begin{corollary} Let $\epsilon >0$. Then, 
with the previous notations, when $\Dcal = \Dcal_N$ for some even value of $N\ge 2$:
\begin{equation} 
	\Pbb[ \{ \exists t \le t_n : x(t) \notin \Xcal \}] \le \epsilon \textrm{ if } 
	n \le \frac{ \frac{\sigma D}{v} + \sqrt{2}\ln(\epsilon/2)}{\sqrt{2}\ln(2)}.
	 \label{cor:exit_spatial_S21D}
\end{equation}
\end{corollary}
\begin{proof}
We use \eqref{eq:proba_exit_before_tn} and 
inequality \eqref{eq:bound_n_exp} from proposition \ref{prop:estimations}. Here $Z_j=a_j \tau_j$, $C=D/v$. We write for $\lambda < \sigma$ and $\Dcal=\Dcal_2$:
\begin{eqnarray}
& \ & 
\E[ e^{\lambda Z_1}] = 
\E[ e^{\lambda a_1 \tau_1}] = 
\frac{1}{2} \E[ e^{\lambda \tau_1}+ e^{-\lambda \tau_1}] =
\nonumber \\ & \ & 
\frac{\sigma}{2} \int_0^\infty e^{-\sigma \tau} (e^{\lambda \tau}+ e^{-\lambda \tau}) d \tau = \frac{\sigma}{2} \left( \frac{1}{\sigma - \lambda} +  \frac{1}{\sigma + \lambda} \right) = \frac{\sigma^2}{\sigma^2-\lambda^2} \ .
\end{eqnarray}
Take now $\lambda = \sigma / \sqrt{2}$. The term in  \eqref{eq:bound_n_exp}
becomes $\frac{\sigma D/(v \sqrt{2}) + \ln(\epsilon/2)}{\ln(2)}$.

For $\Dcal=\Dcal_N$ with $N>2$ (even) denote $d_j=\cos\left( \frac{2 \pi j}{N}\right)$. Then 
\begin{eqnarray} & \ & 
	\E[ e^{\lambda Z_1}] = 
	\E[ e^{\lambda a_1 \tau_1}] = 
	\frac{1}{N} \sum_{j=0}^{N-1} \E[ e^{\lambda d_j \tau_1}] =
	\nonumber \\  & \ & 
	\frac{\sigma}{N} \sum_{j=0}^{N-1} \int_0^\infty e^{-\sigma \tau} e^{\lambda d_j \tau} d \tau =\frac{1}{N} \sum_{j=0}^{N-1} \frac{\sigma}{\sigma - d_j\lambda}.
\end{eqnarray}
In the sum over $j$, for even values of $N$, values can be regrouped by 
associating some value $d$ with the value $-d$. There are $N/2$ such couples and each of them will contribute 
$ \frac{\sigma}{\sigma - d\lambda} + \frac{\sigma}{\sigma  + d\lambda} =
\frac{\sigma^2}{\sigma^2  - d^2\lambda^2} \le \frac{\sigma^2}{\sigma^2  - \lambda^2}
$. We conclude as in the case $N=2$.

\end{proof}
This corollary can be used to know how many elementary collisions can be aggregated while keeping the chances to reach the frontier very small. In practice we will improve this bound but it has the merit to show that $\sigma D/v$ should be large enough with respect to $\ln(\epsilon)$.
Conservative values for $\epsilon$ will be of order $10^{-9}$ which is a chance in a billion to be wrong and assume that particle still stays inside the domain when in reality it has exited. This would give
$ \sqrt{2}\ln(\epsilon/2) =-30.28$ which leads, using $ \sqrt{2}\ln(\epsilon/2) \sim 0.98<1$ to:
\begin{equation}
n \le \frac{\sigma D}{v} - 31. \label{eq:decision_rule_n_spatial_notnormal}
\end{equation}

This is a more useful value 
than  $  \frac{2}{N} \cdot 10^{-9} \cdot \left( \frac{\sigma D}{v} \right)^2$ from remark~\ref{rem:markov}. Of course, which one is larger depends on the precise value of $\frac{\sigma D}{v}$ but in the regimes where this is of interest to us  $\frac{\sigma D}{v}$  can be quite large. From a qualitative point of view, combining the two behaviors is even better; the goal would be to have an estimate that contains 
$\left( \frac{\sigma D}{v} \right)^2$ with a very weak dependence on $\epsilon$, for instance logarithmic or even weaker. This will be provided in the context of the normal approximation presented in section \ref{sec:normal_spatial} and using Hoeffding inequality in the next result. 
\begin{proposition}
Let $\epsilon >0$. Then, 
with the previous notations, for even values of $N$:
\begin{equation} 
\Pbb[ \{ \exists t \le t_n : x(t) \notin \Xcal \}] \le \epsilon \textrm{ if } 
2 n \left[\ln(n)+ \ln\left(\frac{2}{\epsilon}\right)\right]^2  \ln\left(\frac{8}{\epsilon}\right) \le \frac{ D^2 \sigma^2}{v^2} 
\label{eq:exit_spatial_hoeffding}
\end{equation}
\label{cor:exit_spatial_hoeffding}
\end{proposition}
\begin{proof}
We invoke~\eqref{eq:reflection_ineq} from proposition \ref{prop:estimations} for $Z_j=a_j \tau_j$, $C=D/v$. We choose some level to be defined $\xi >0$ and recall that for $U\simeq Exp(1)$: $\Pbb[U \ge \xi] = e^{-\xi}$ and thus 
 $\Pbb[\tau_i \ge \xi/\sigma] = e^{-\xi}$. We can replace each $\tau_i$ by 
   $\tilde{\tau_i} = \min\{ \tau_i, \xi/\sigma \}$. 
Note that $\Pbb[\tilde{\tau_i} = \tau_i]=1-e^{-\xi}$ and 
 $\Pbb[\tilde{\tau_i} = \tau_i, \ \forall i \le n ]\ge 1-n e^{-\xi}$.

Consider the modified trajectory $\tilde{x}(t)$ constructed as $x(t)$ but with $\tilde{\tau_i}$ instead of ${\tau_i}$. Previous estimation informs that 
$\Pbb[ \exists t \le t_n : \tilde{x}(t) \neq x(t)]\le n e^{-\xi}$.
We will use the following  Hoeffding inequality:
if $\chi_1$, ..., $\chi_n$ are i.i.d. random variables such that 
$\alpha_i \le \chi_i \le \beta_i$ almost surely, then
$\Pbb[ | \sum_{i=1}^n (\chi_i - \E[\chi_i]) | \ge \mathfrak{m} ]
\le 2\exp \left(-{\frac {2 \mathfrak{m}^{2}}{\sum_{i=1}^{n}(\beta_{i}-\alpha_{i})^{2}}}\right)$.
We use this Hoeffding inequality for the variables $\tilde{Z_j}=a_j \tilde{\tau_j}$ which have values inside $[-\xi/\sigma,\xi/\sigma]$ and average to zero:
\begin{eqnarray}
	& \ & 
	\Pbb[ \{ \exists t \le t_n : x(t) \notin \Xcal \}] \le
n e^{-\xi}+ \Pbb[ \{ \exists t \le t_n : \tilde{x}(t) \notin \Xcal \}]  
	\nonumber \\  & \ & 
\le n e^{-\xi}+
2 \Pbb \left[ \left|\sum_{j=1}^n \tilde{Z_j}\right| \ge  \frac{D}{v}\right] 
\nonumber \\ & \ & 
\le  n e^{-\xi}+ 4 e^{- \frac{2 D^2 \sigma^2}{4 n v^2 \xi^2}}.
\end{eqnarray}

We now choose $\xi$ such that $n e^{-\xi}=\epsilon/2$ i.e.
$\xi=\ln(n)+ \ln(2/\epsilon)$. Then, it is enough to find $n$ such that 
$4 e^{- \frac{2 D^2 \sigma^2}{4 n v^2 \xi^2}} \le \epsilon/2$ to conclude.
Replacing the value of $\xi$ we obtain the required estimation.
\end{proof}
This result is to be used when $\frac{D\sigma}{v}$ is large. For example, the bound in equation \eqref{eq:decision_rule_n_spatial_notnormal} is better for values of $\frac{D\sigma}{v}$ up to $\simeq 5\times 10^4$. On the contrary, for 
 $\frac{D\sigma}{v}\simeq 10^{6}$, estimation 
 \eqref{eq:exit_spatial_hoeffding} gives a "$n$" one order of magnitude larger than that of \eqref{eq:decision_rule_n_spatial_notnormal}; for $\frac{D\sigma}{v}\simeq 10^{7}$ more than two orders of magnitude are gained.

\subsection{Spatial boundary treatment, normal approximation} \label{sec:normal_spatial}

We consider now the normal approximation. Then
in this case $\sum_{j=1}^n a_j \tau_j$ looks like a 
Brownian motion and, in view of the estimation
 \eqref{eq:reflection_ineq} if we accept and error $\epsilon$,
 we will set $n$ such that:  
\begin{equation}
2 \Pbb \left[ \left|\sum_{j=1}^n a_j \tau_j\right|  \ge D/v \right] \le \epsilon.
\label{eq:proba_exit_before_tn_normal}
\end{equation}
But $\sum_{j=1}^n a_j \tau_j$ has mean $0$ and variance $n \varsigma^2 / \sigma^2$
so the probability to be larger than $D/v$ will be small when 
$ \frac{D \sigma }{v \varsigma \sqrt{n}}$ is large enough (of the order of $6$), giving the aggregation rule:
\begin{equation}
n \le \frac{1}{50} \left(\frac{D \sigma}{v \varsigma}\right)^2.
\label{eq:decision_rule_n_space_normal}
\end{equation}
As before, this rule will be used when $n$ is large enough, we take 
$n$ larger than some $n_\Ncal$ that is set to $n_\Ncal:=300$. Note that the dependence with respect to $D\sigma$ is quadratic and not linear as in corollary~\ref{cor:exit_spatial_S21D} (recall that  $\varsigma=1$ when $\Dcal=\{-1,1\}$). The dependence on the tolerance $\epsilon$ is very weak, because the $1-\epsilon/2$ quantile of the normal law increase very slowly when $\epsilon \to 0$.

\subsection{Summary of the aggregation rules}
We summarize the estimations obtained in table~\ref{table:estim_n}.

{
\begin{table}[h]
\begin{tabular}{@{}l|l|l|l|l}
\toprule
situation & time  & condition&  spatial  & condition\\
\midrule
normal   & eq.~\eqref{eq:normal_approx_tn}:	
& $n \ge n_\Ncal:=300$   & eq.~\eqref{eq:decision_rule_n_space_normal}: $n \le \frac{1}{25 N} \left(\frac{D \sigma}{v}\right)^2$  & $n \ge n_\Ncal:=300$  \\
approximation  & $n \le T\sigma +18 - 6\sqrt{T\sigma+9}$&(empirical)   &  & (empirical)\\
\midrule
conservative &  eq.~\eqref{eq:tail_estimates_tn}:	$n \le T\sigma /2$
  & $n\ge 50$  & eq.~\eqref{cor:exit_spatial_S21D} \&
\eqref{eq:decision_rule_n_spatial_notnormal}: $n \le \frac{\sigma D}{v} - 31$
 & $n\ge 1$ \\
(probability &  &  & eq.~\eqref{eq:exit_spatial_hoeffding}:
$n \left[\ln(n)+ 21.42\right]^2 $
&  \\
 $\ge 1- 10^{-9}$) &  &  & $\le \frac{1}{45.61}\frac{ D^2 \sigma^2}{v^2}$&  \\
\midrule
\end{tabular}
\caption{Summary of estimations of the number of steps. We used that $\varsigma^2=N/2$ for $\Dcal=\Dcal_N$.}\label{table:estim_n}%
\end{table}
}

\subsection{Sampling once aggregation level $n$ is given}

Previous considerations helped to choose a $n$ such that $n$ collisions later the particle has a very high probability to be still in the space-time domain $\Xcal \times [0,T]$. But we still need to place the particle somewhere i.e., we need to explain how to sample from the distribution of the position and time coordinates of the particle after $n$ collision steps.

As soon as $n>1$ the direction of the particle is very easy to sample: it is 
a value from $\Dcal$ sampled uniformly. For the time $t_n$ we take a gamma variable with shape $n$ and scale $1/\sigma$ (average = $n/\sigma$). What about the position $x(t_n)$ ? We know that 
$x(t_n)= x_0 +  \sum_{j=1}^n a_j \tau_j$. To compute this efficiently we sample from the multinomial distribution $\Mcal(1/N,...,1/N)$ with $N$ events and obtain positive integers $n_j$ that sum up to $n$ with the convention that $n_j$ will represent the number of times the value $d_j\in \Dcal$ has been taken by some $a_\ell$. So finally the term $ \sum_{j=1}^n a_j \tau_j$ will be a sum of $N$ gamma distributions, with $j$-th term being of parameters $n_j$ and $d_j/\sigma$. The resulting procedure is described in the algorithm~\ref{algo:accelerated}.

\begin{algorithm}
\caption{Aggregating collision escape sampling algorithm, normal approximation}
\label{algo:accelerated}
\noindent {\bf Inputs : }  no. directions $N$, spatial size $L$, initial position $x_0$, total time $T_f$, collision parameter $\sigma$, speed $v$, approximation type  $apr \in \{ ``normal", ``not \ normal" \}$.

\noindent {\bf Outputs : }  space-time exit point $x^\star, a^\star, t^\star$, number of collisions $c$
\begin{algorithmic}[1]
\Procedure{Single collision}{$N,L,x,t,T_f,\sigma,v$}
\State set $\Xcal=[0,L]$
\State sample $\tau \simeq Exp(1/\sigma)$, $a$ uniform from $\Dcal$
\If{$t+\tau < T_f$ and $x+v a \tau \in \Xcal$}
\State return $(t+\tau,x+v a \tau,a ,\text{`not escaped'})$
\Else 
\State compute time to escape when either $T_f$, or $\partial \Xcal$ are reached: 
$\Delta t^\star = \min\{\tau, T_f-t,\frac{dist(x,\partial \Xcal)}{va}\}$ 
\State  return $(t+\Delta t^\star,x+v a \Delta t^\star,a,\text{`escaped'})$
\EndIf
\EndProcedure

\Procedure{escape time sampling}{$N,L,x_0,T_t,\sigma,v$}
\State Set $t=0$, $x=x_0$, collision counter $c=0$, flag=`not escaped', 
\If{apr==``normal"} \ $n_\Ncal=300$
\Else \ $n_\Ncal=50$
\EndIf
\State compute $d_k = \cos(2 \pi j/N)$, $j=0,..., N-1$.
\While{flag=`not escaped'}
\State set $D := \min\{x,L-x\}$, $T=T_t-t$.

\If{apr==``normal"} \ $n_c=\min \left\{ T\sigma +18 - 6\sqrt{T\sigma+9}, \frac{1}{25 N} \left(\frac{D \sigma}{v \varsigma}\right)^2 \right\}$ 
\Else \  $n_c=\max \left\{ n \le T\sigma/2 \left| 
n \le \frac{\sigma D}{2}-31 \text{ or }
n [\ln(n)+21.42]^2 \le \frac{1}{45.61}  \frac{D^2 \sigma^2}{v^2} \right. \right\}$ 
\EndIf
\If{$n_c\ge n_\Ncal$}:
\State sample $\tau \simeq \Gamma(n,1)$, $n_1,...,n_N \simeq \Mcal(1/N,...,1/N)$,
  $\tau_\ell \simeq \Gamma(n_\ell,1)$, $\ell=1, ...,N$
\If{$t+\tau/\sigma>T$ or $x+\sum_{\ell=1}^N d_\ell \tau_\ell/\sigma \notin ]0,L[$} 
\State print `error in' apr ` approximation' 
\State $(t,x,a,flag)=\text{SINGLE COLLISION}(N,L,x,t,t+T,\sigma,v)$.
\State $c=c+1$
\Else
\State set $t=t+\tau/\sigma$, $x=x+\sum_{\ell=1}^N d_\ell \tau_\ell/\sigma$
\State $c=c+n_c$
\EndIf
\Else 
\State $(t,x,a,flag)=\text{SINGLE COLLISION}(N,L,x,t,t+T,\sigma,v)$.
\State $c=c+1$
\EndIf
\EndWhile 
\State return $(t,x,a)$, $c$
\EndProcedure
\end{algorithmic}
\end{algorithm}

\section{Numerical results} \label{sec:numerical}

\subsection{One dimension two directions} \label{sec:onedstwo}

We start with the 1D case where the number of directions is $N=2$, i.e. the $S_2$ model. This model is interesting in itself and not necessarily as a discretization of the situation when $\Dcal$ is the unit sphere (see the introduction).

\begin{figure}[htbp!]
	\includegraphics[width=0.7\textwidth]{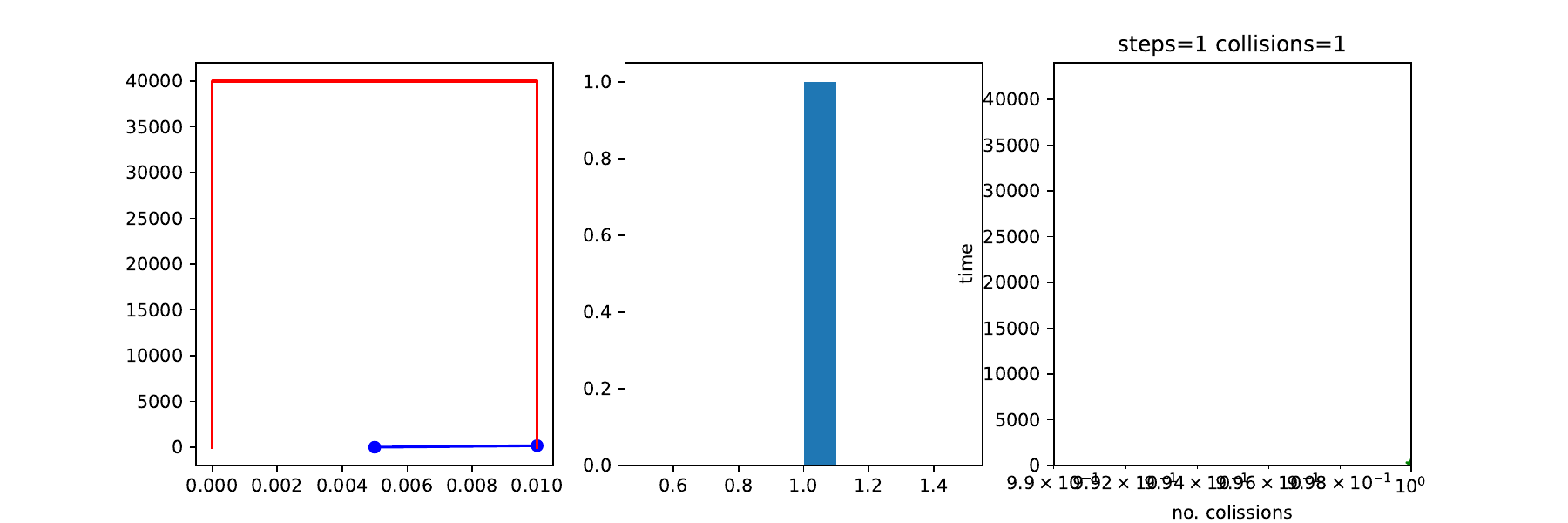}
	
	\includegraphics[width=0.7\textwidth]{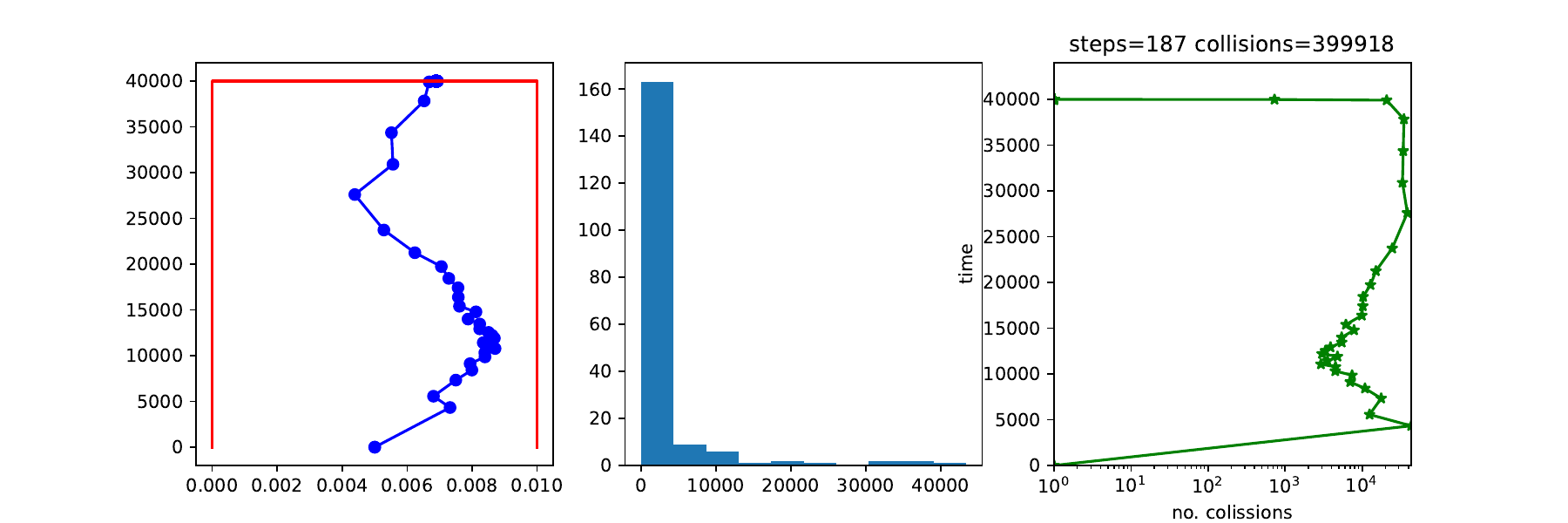}
	
	\includegraphics[width=0.7\textwidth]{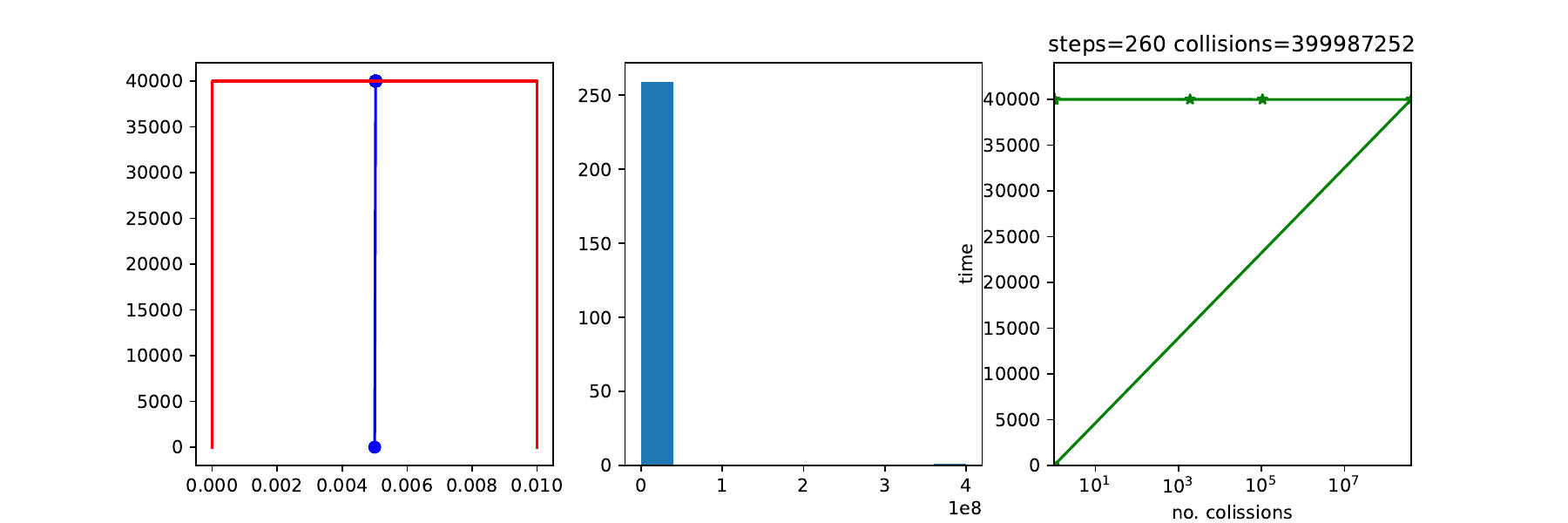}
	
	\includegraphics[width=0.7\textwidth]{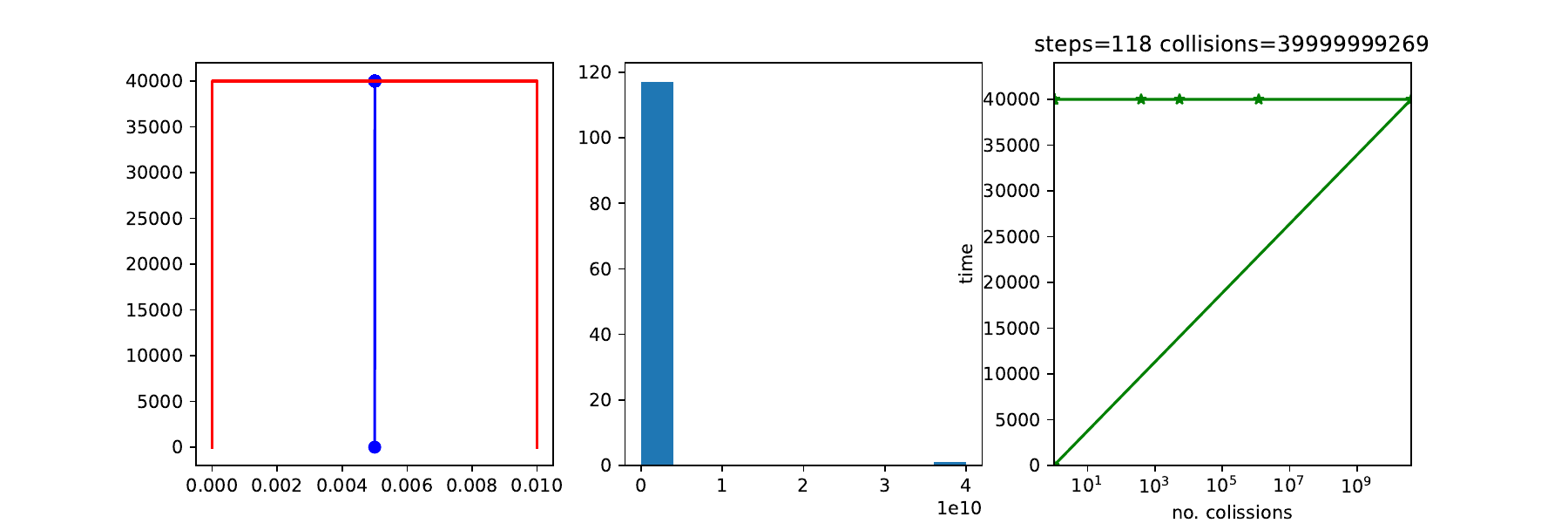}
	\caption{\small Examples of trajectories for the 1D $S_2$ test case with $L=0.01cm$, 
 $T_f=4\times 10^4fs$, $v=3.0\cdot 10^{-5} fs/ cm$, $\sigma \in \{10^{-2}$, $10$,  $10^{4}$, $10^{6}\}$ ($cm^{-1}$), $x_0=L/2$, initial direction $+1$ (oriented towards right).
 Each row is a 
 single trajectory, each trajectory corresponding to a  
 different value of $\sigma$. First column plots the particle evolution $x(t)$ with ordinate axis being the time and abscissas the location in the $[0,L]$ segment. Second column displays the histogram of the $n_c$ values (cf algorithm \ref{algo:accelerated}.) The third column gives information on the total steps taken and total number of collisions treated. The abscissas are the number of collisions and the ordinate axis is time in femtoseconds. 
 The particle in the first row has an exit direction $a^\star=+1$ because no collision takes place. The particles in the last three rows have a exit direction $a^\star$ that will be drawn at random from $\{-1,1\}$ because in all cases exit is due to the fact that time was consumed.  
}\label{fig:1D_traj}
\end{figure}

\begin{figure}[htbp!]
	\includegraphics[width=0.7\textwidth]{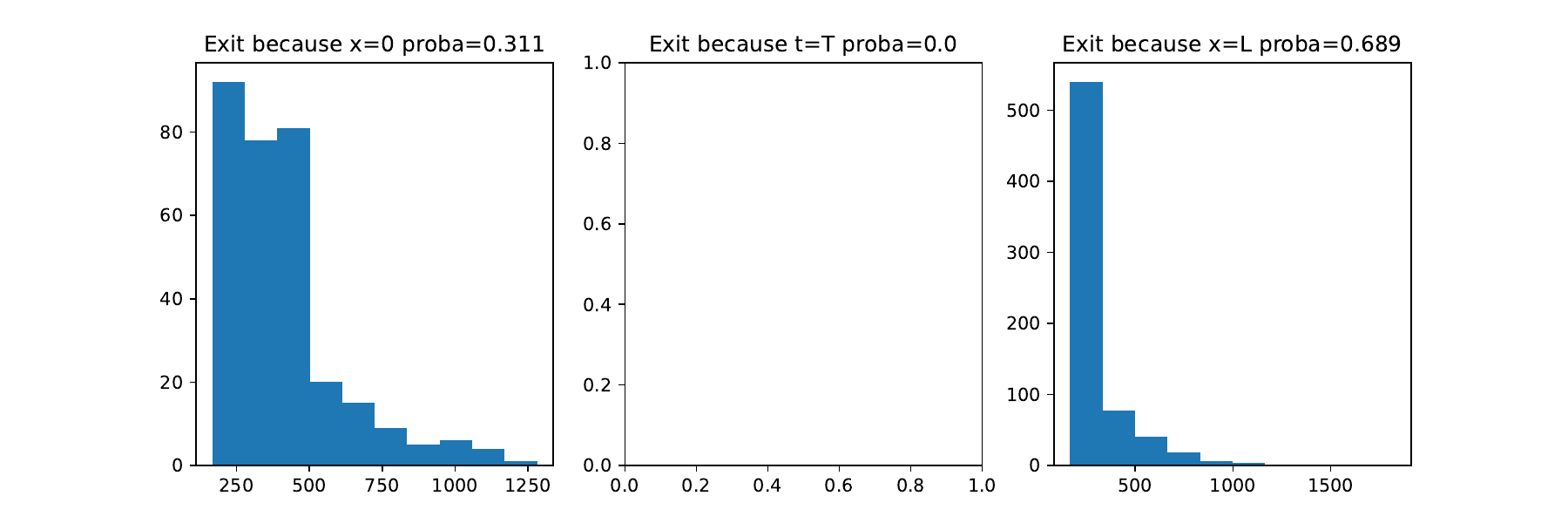}

	\includegraphics[width=0.7\textwidth]{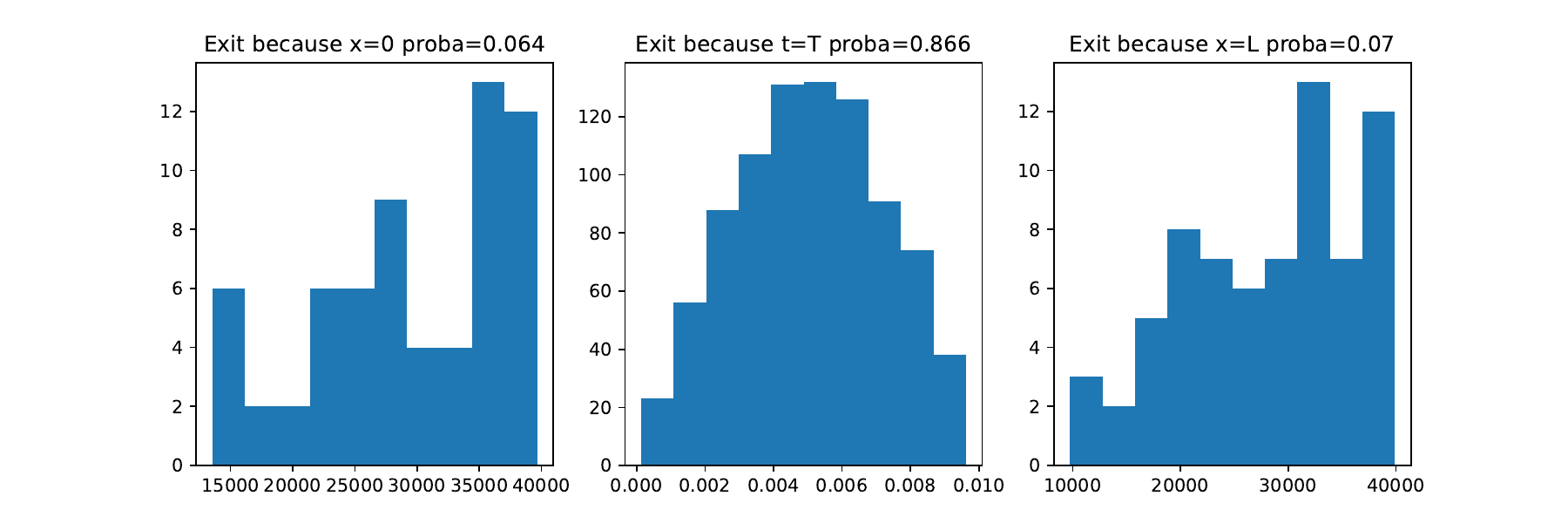}
	
	\includegraphics[width=0.7\textwidth]{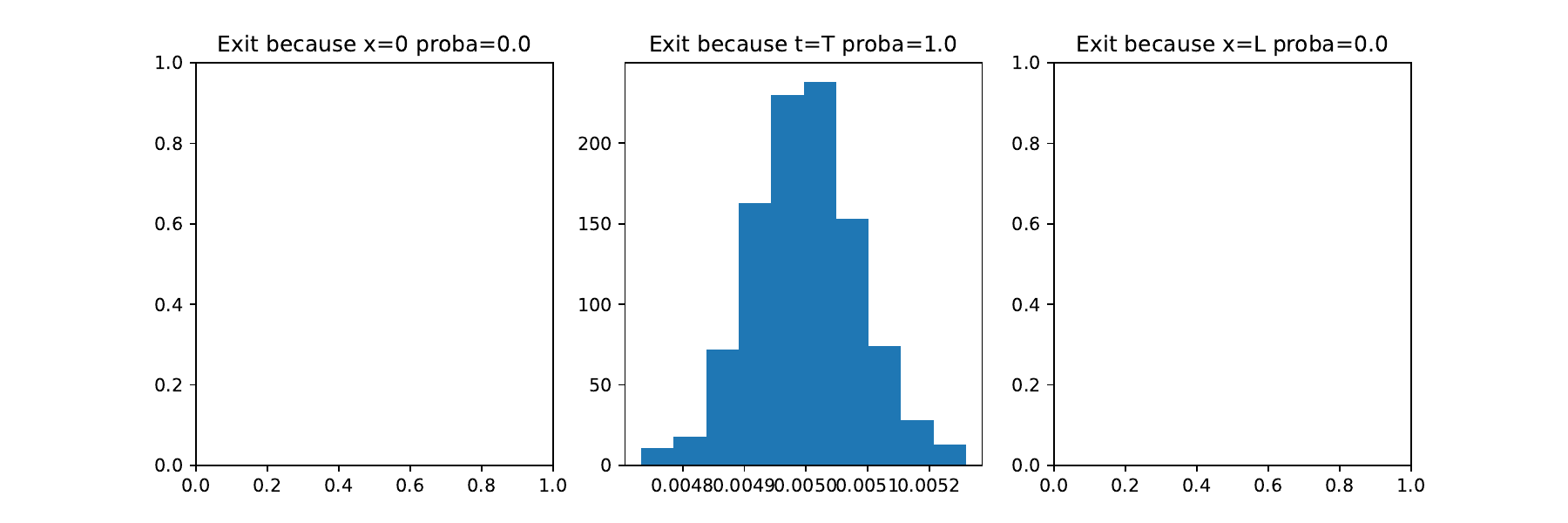}
	
	\includegraphics[width=0.7\textwidth]{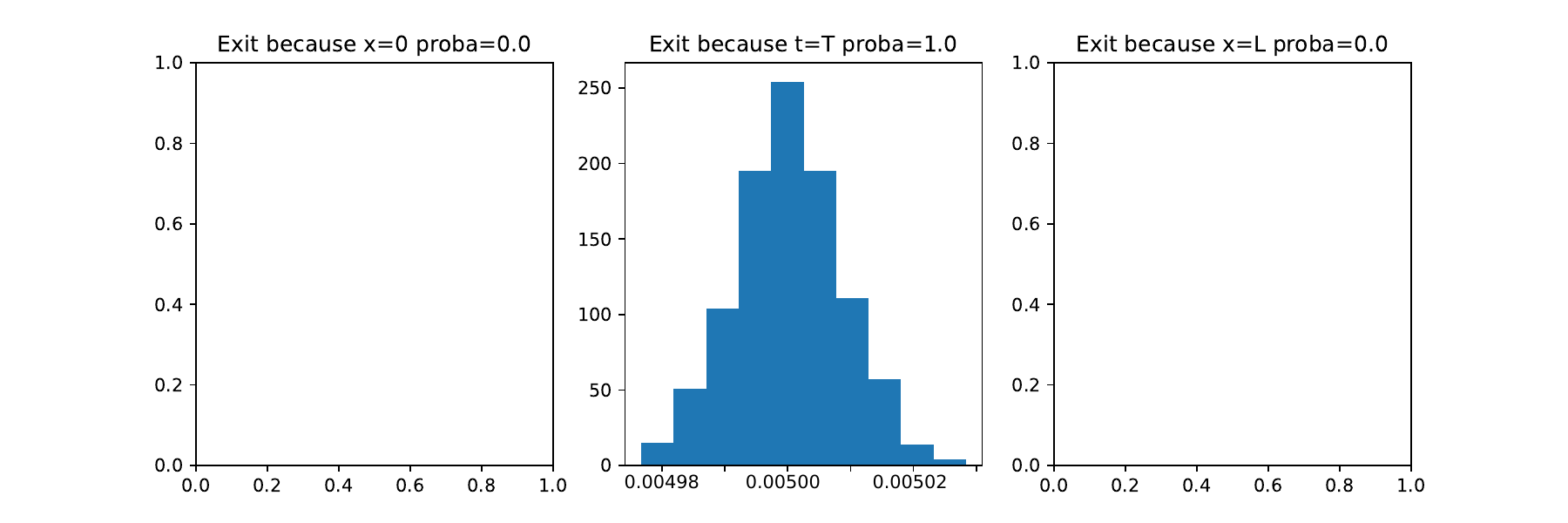}
	\caption{The escape laws corresponding to figure~\ref{fig:1D_traj}. Same convention for the signification of the rows is used.
For the middle column the abscissas are time values expressed in femtoseconds; for all other plots the abscissas are $x$ values in the segment $[0,L]$. 
 Each column is a conditional distribution: in the first column 
 are drawn histograms of $t^\star$ for 
  escape points with $x(t^\star)=0$ (escape through the left extremity of the domain), second column 
  draws histograms of $x^\star=x(T)$ for escape points with
  with $t=T$ (escape because time is consumed) and third histograms of $t^\star$ for 
  escape points with $x(t^\star)=L$
   (escape through the right extremity of the domain). The probability of escape for each alternative 
   is given above the plot.  When a histogram is void there is no escape for the alternative the histogram is supposed to represent. For instance, in the third row, first column one expects to see the histogram of escape time $t^\star$ conditioned by the fact that escape occurred through the left side i.e. $x(t^\star)=0$ (before time $T$ and before reaching the right side). But this never happens because in this case $\sigma$ is too large, many collisions occur and the particle does not have the time to reach the left side (the right side neither in fact) before time $T$.}
	\label{fig:1D_laws}
\end{figure}

For the numerical tests, we set the segment length to $L=0.01$ cm, final time to $T=4\times 10^4$ fs, speed $v=3.0\cdot 10^{-5} fs/ cm$, and  $\sigma \in \{10^{-2}$, $10$,  $10^{4}$, $10^{6}\}$ (unit is $cm^{-1}$). 
The values have been chosen to see all regimes (see also results below): for $\sigma=10^{-2}$ there are very few collisions while for $10^6$ there are a large number of them. Then the sampling inside these bounds is to obtain an almost log-uniform $T\sigma$ grid (mean number of collisions).
The initial point is in the middle of the spatial interval ($x_0=L/2$) thus $D=L/2$.
We plot  in figure~\ref{fig:1D_traj} some examples of trajectories and in figure~\ref{fig:1D_laws} the resulting escape laws.
The values of the parameters $T\sigma$ and $D \sigma/v$ at the initial time are given in
table~\ref{table:valuestsigmaetc}.

\begin{table}[htbp!]
\begin{tabular}{@{}l|l|l|l|l}
\toprule
& $\sigma=10^{-2}$ &  $\sigma=10^{1}$&  $\sigma=10^{4}$  & $\sigma=10^{6}$\\
\midrule
$T \sigma$ & $4\times 10^{2}$  &  $4\times 10^{5}$ &  $4\times 10^{7}$  & $4\times 10^{10}$\\
\midrule
$D\sigma/v$&  $\frac{10}{6}$  & $\frac{10^4}{6}$&   $\frac{10^7}{6}$    & $\frac{10^{11}}{6}$ \\
\midrule
\end{tabular}
	\caption{Values of $T\sigma$ and $D\sigma/v$ for simulations in section~\ref{sec:onedstwo}.}\label{table:valuestsigmaetc}%
\end{table}

For   $\sigma=10^{-2}$ (first row of figures~\ref{fig:1D_traj} and \ref{fig:1D_laws}) the particle has very few collisions (here $2$) and exists the spatial domain well in advance of the final time $T_f$. The aggregation mechanism is never activated. The escape distributions concentrate on the borders $x=0$ and $x=L$; the middle column depicts 
the escape law conditional to having escaped because final time is reached; here this column has no mass at all. 

For $\sigma=10^{1}$ (second row of figures~\ref{fig:1D_traj} and~\ref{fig:1D_laws}) the computation can be accelerated by aggregating several collisions, up to $\simeq 4\times 10^4$ for an average of 
$399918/187=2138$ collisions per step. The particular trajectory depicted here exists because final time is reached but $\simeq 14\%$ of trajectories exit because the spatial border is touched first.

For $\sigma=10^{4}$ and $\sigma=10^{6}$ (last two rows of figures~\ref{fig:1D_traj} and~\ref{fig:1D_laws}) the  particle undergoes an important number of collisions. The proposed procedure turns out to be very useful with a number of collisions per aggregated step being 
$\frac{399987252}{266}=1.503 \times 10^6$ and $\frac{39999999269}{118}=3.390 \times 10^{8}$ respectively. 
The procedure reaches thus acceleration factors of up to $10^8$ and the numerical resolution is very expensive without it.
All particles exit because the time is up, none exists through the spatial borders: the particle behavior is that of a random walk around the initial point.

\subsection{One dimension, $N$ directions} \label{sec:onedsn}

We take now $N >2$ directions, i.e., the $S_N$ model. Such a model can for instance be used as a discretization of the situation when $\Dcal$ is the whole unit sphere. We will not test the same things as before but instead investigate the two possible sources of error: the fact that $N$ is finite (so $\Dcal_N$ is not a perfect representation of the unit sphere)
and the fact that we used the normal approximation. Moreover we will take the most difficult test case which is the situation of a long time and a moderate value of $\sigma$, neither in the ballistic (small $\sigma$) nor in the diffusion (large $\sigma$) regime: $T=2 \times 10^4$fs, $\sigma=1$. The initial point is in the middle of the spatial interval ($x_0=L/2$).

\begin{figure}[htbp!]
	\includegraphics[width=0.7\textwidth]{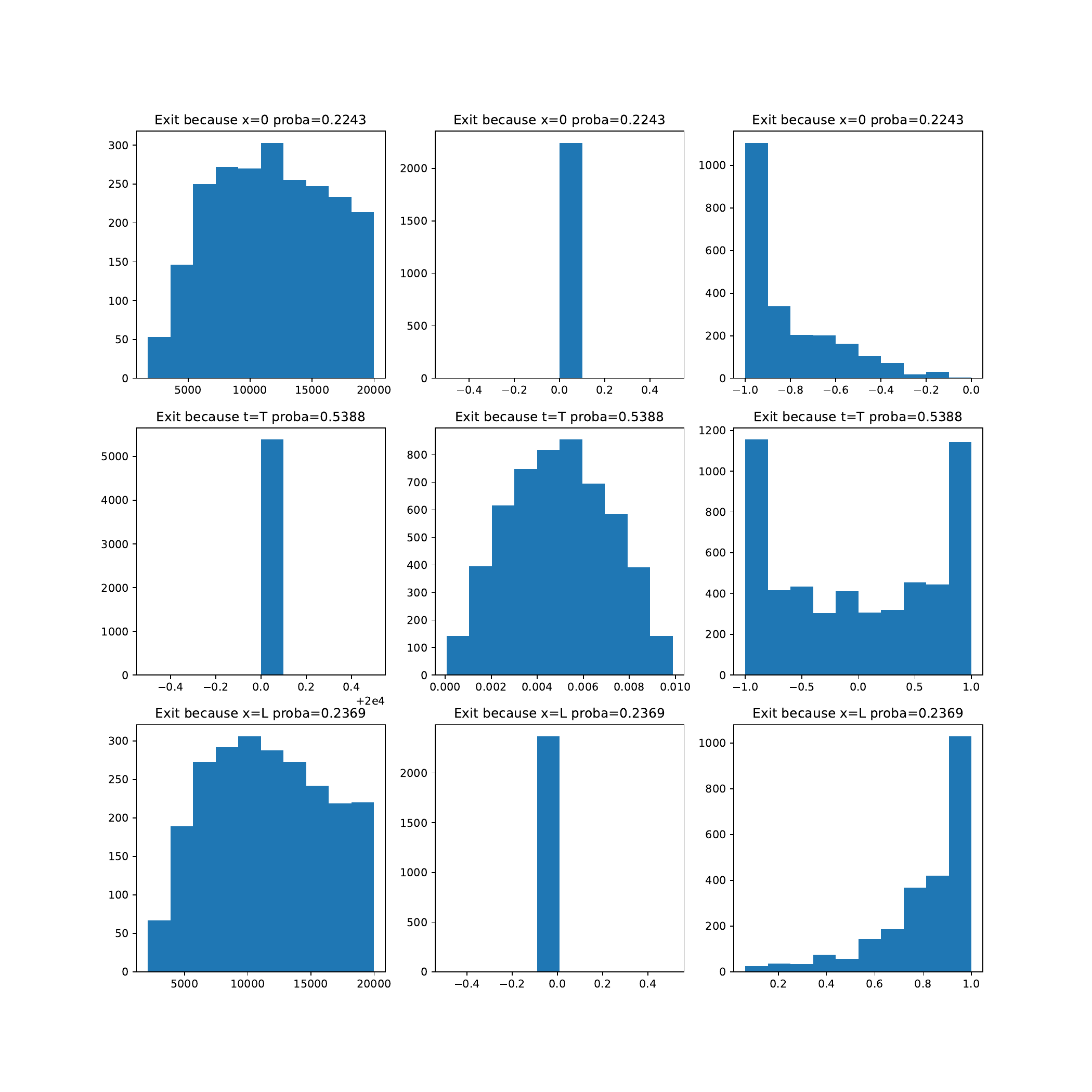}
	\caption{Results for the setting in section~\ref{sec:onedsn} for $n_\Ncal=300$, $N=100$. The histograms of the escape distributions $t,x,a$ are plotted as follows: first column are histograms of the time values $t^\star$, second column the positions $x^\star$ and third the directions $a^\star$. Each row is a conditional distribution: the first row are escape points with $x^\star(t^\star)=0$ (escape through left side), second row with $t^\star=T$ (escape due to total time $T$ being consumed) and third with $x^\star(t^\star)=L$ (escape through the right side).}
	\label{fig:1D_sn_laws}
\end{figure}

\begin{figure}[htbp!]
	\includegraphics[width=0.7\textwidth]{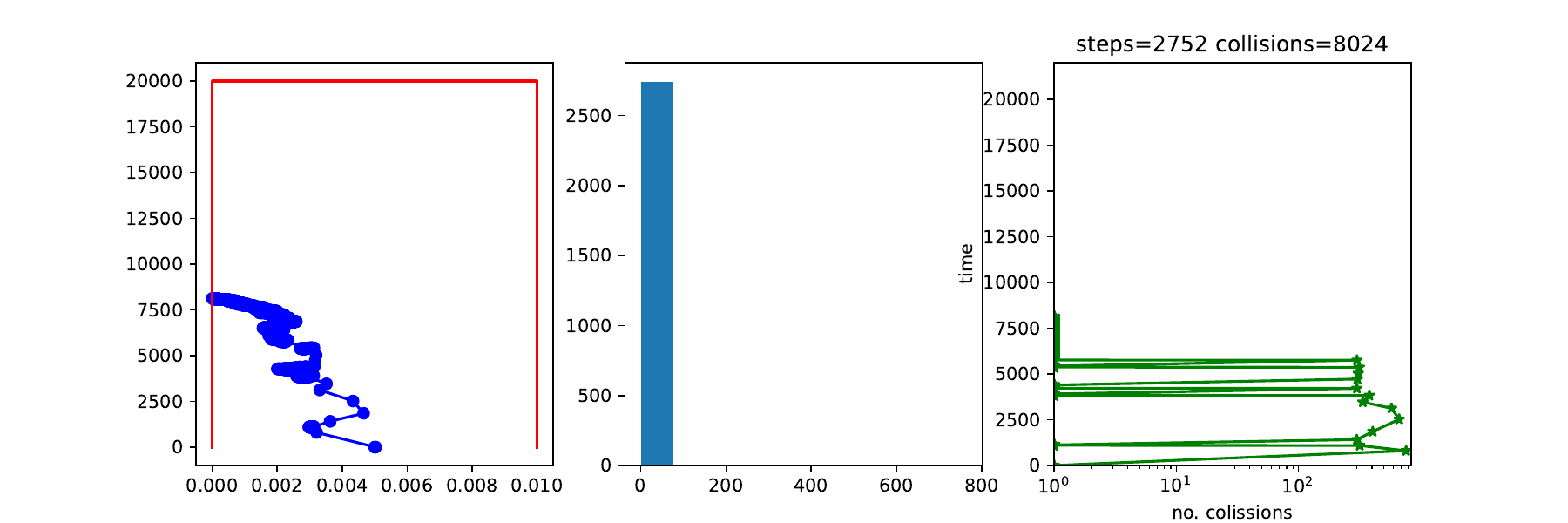}
	\caption{One example trajectory corresponding to results in figure	\ref{fig:1D_sn_laws}, i.e. $n_\Ncal=300$, $N=100$. {\bf Left plot:} the trajectory $x(t)$ of the particle, $y$ axis is the time. Exit direction is $a^\star=-1$. {\bf Middle plot:} the distribution of the number of collisions that have been aggregated. {\bf Right plot:} the number of collisions with time in the $y$ axis. Many are equal to $1$ but some go as high as $600-800$.}		
	\label{fig:1D_sn_traj}
\end{figure}

We consider several test cases: parameters $N=10$ can be or $N=100$; the $n_\Ncal$ parameter (that decides when the normal approximation is to be used) can be $n_\Ncal=50$, $n_\Ncal=300$ or $n_\Ncal=1000$. In each case we compute $10^4$ escape points $t,x,a$. The nominal values are : $n_\Ncal=300$ cf. table~\ref{table:valuestsigmaetc} and $N=100$. The results for these nominal parameters are given in figures~\ref{fig:1D_sn_laws} and~\ref{fig:1D_sn_traj}.  
It is seen in figure~\ref{fig:1D_sn_laws} that in this case all escape sides are populated, i.e., the particle can escape either through $x=0$ or $t=T$ or $x=L$. As expected the situation is symmetric and this is confirmed by the fact that the probability to exit though $x=0$ is the same as that for $x=L$ (up to precision $10^{-2}$).
The figure~\ref{fig:1D_sn_traj} presents an example of trajectory. It is seen that the aggregation procedure has been effective because it reduced the total number of steps as can be seen comparing the total number of collisions with the actual number of steps taken. As expected, when the trajectory (the leftmost plot) approaches the boundary the collisions are treated one by one and when the trajectory is close to the middle of the interval circa $600$ collisions are aggregated together.

\begin{figure}[htbp!]
	\includegraphics[width=0.7\textwidth]{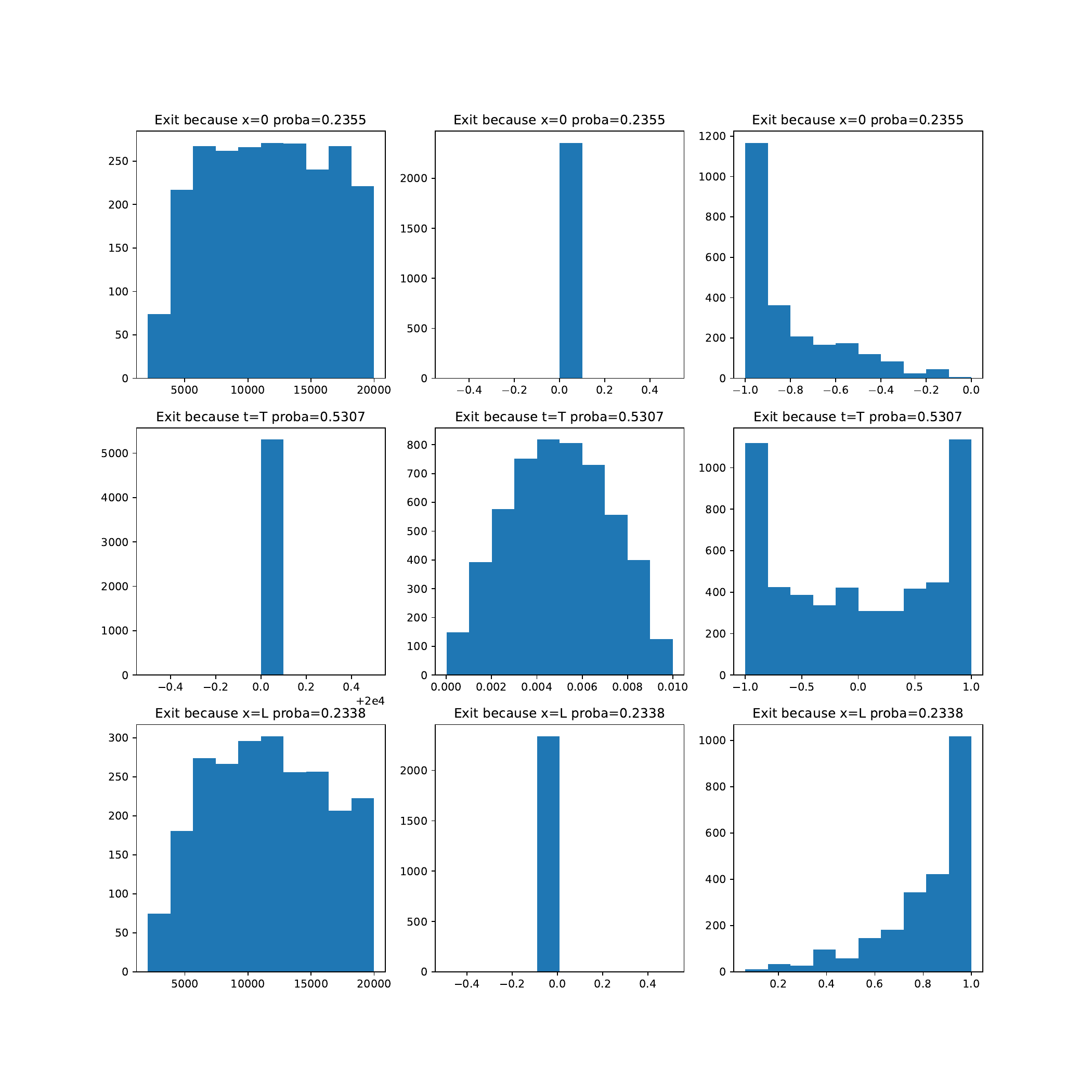}
	\caption{Results for the setting in section~\ref{sec:onedsn} for $n_\Ncal=1000$, $N=100$.}
	\label{fig:1D_sn_laws_100_1000}
\end{figure}

\begin{figure}[htbp!]
	\includegraphics[width=0.7\textwidth]{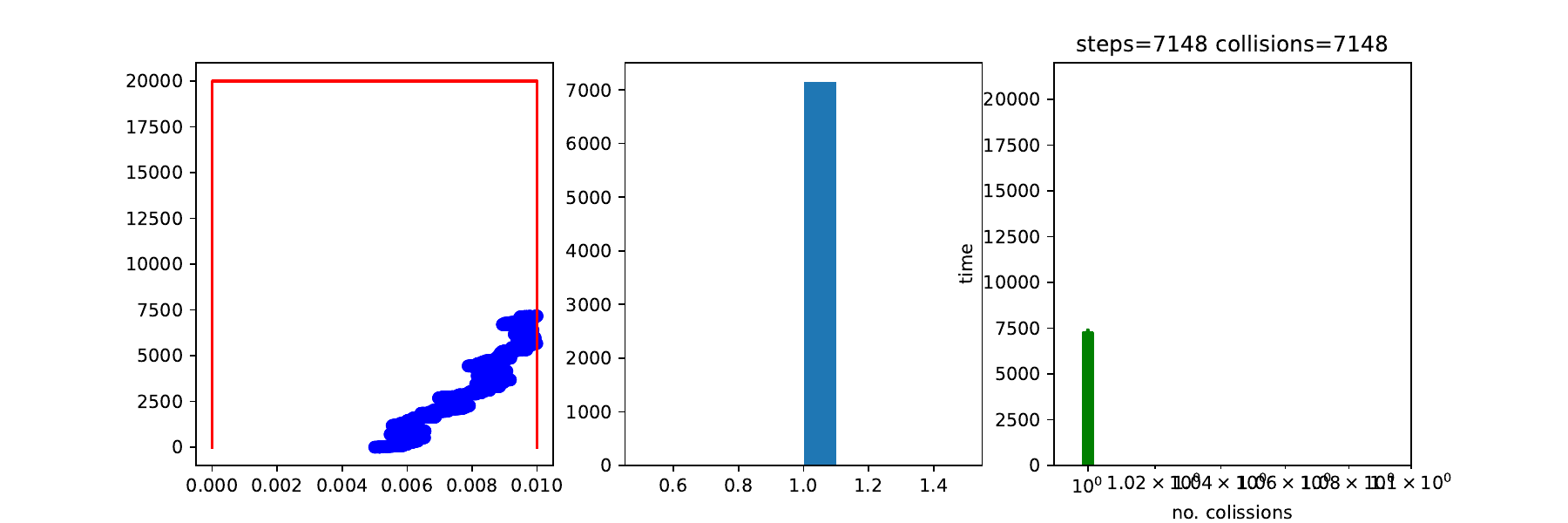}
	\caption{One example trajectory corresponding to results in figure	\ref{fig:1D_sn_laws_100_1000}, i.e., $n_\Ncal=1000$, $N=100$. Signification of the plots is as in figure \ref{fig:1D_sn_traj}.}
	\label{fig:1D_sn_traj_100_1000}
\end{figure}

We make now variations around the nominal parameters. The results for $n_\Ncal=1000$ and $N=100$ are given in figures~\ref{fig:1D_sn_laws_100_1000}
and~\ref{fig:1D_sn_traj_100_1000}. Here $n_\Ncal$ is large enough so that all collisions are treated alone, no aggregation is used so the result is the one that we could have obtained without any aggregation procedure, cf. figure~\ref{fig:1D_sn_traj_100_1000} middle where the histogram is a Dirac mass in $1$. On the other hand figure~\ref{fig:1D_sn_laws_100_1000} shows that the 
escape distributions are very similar to that in figure \ref{fig:1D_sn_laws} which shows that the aggregation procedure obtains comparable quality at lower costs. 
We have also tested $n_\Ncal=50$ and $N=100$ and the results (not shown here) are the same as in~\ref{fig:1D_sn_laws} showing that our estimate $n_\Ncal=300$ is actually a conservative one.

\begin{figure}[htbp!]
	\includegraphics[width=0.7\textwidth]{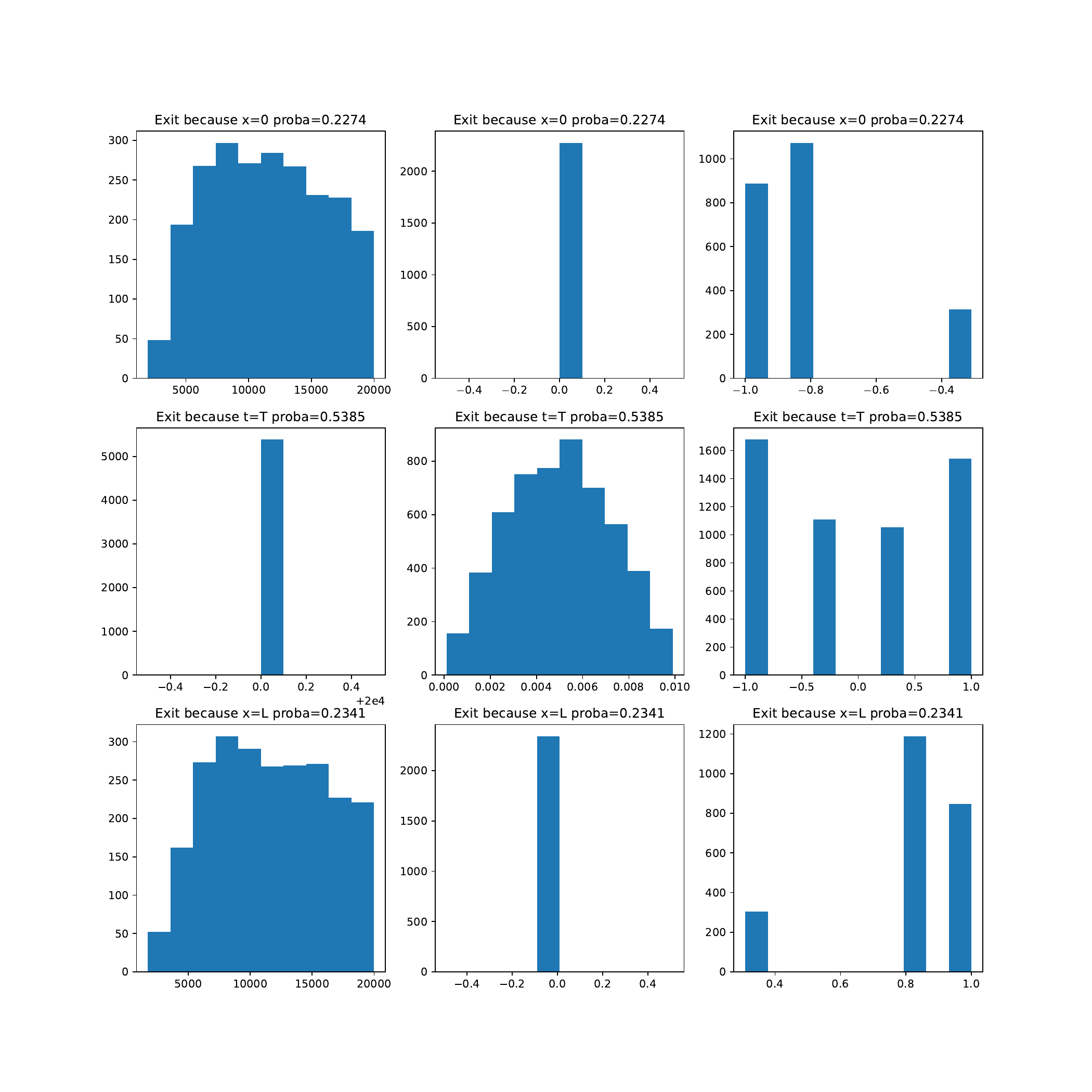}
	\caption{Results for the setting in section~\ref{sec:onedsn} for $n_\Ncal=300$, $N=10$.}
	\label{fig:1D_sn_laws_10_300}
\end{figure}

\begin{figure}[htbp!]
	\includegraphics[width=0.7\textwidth]{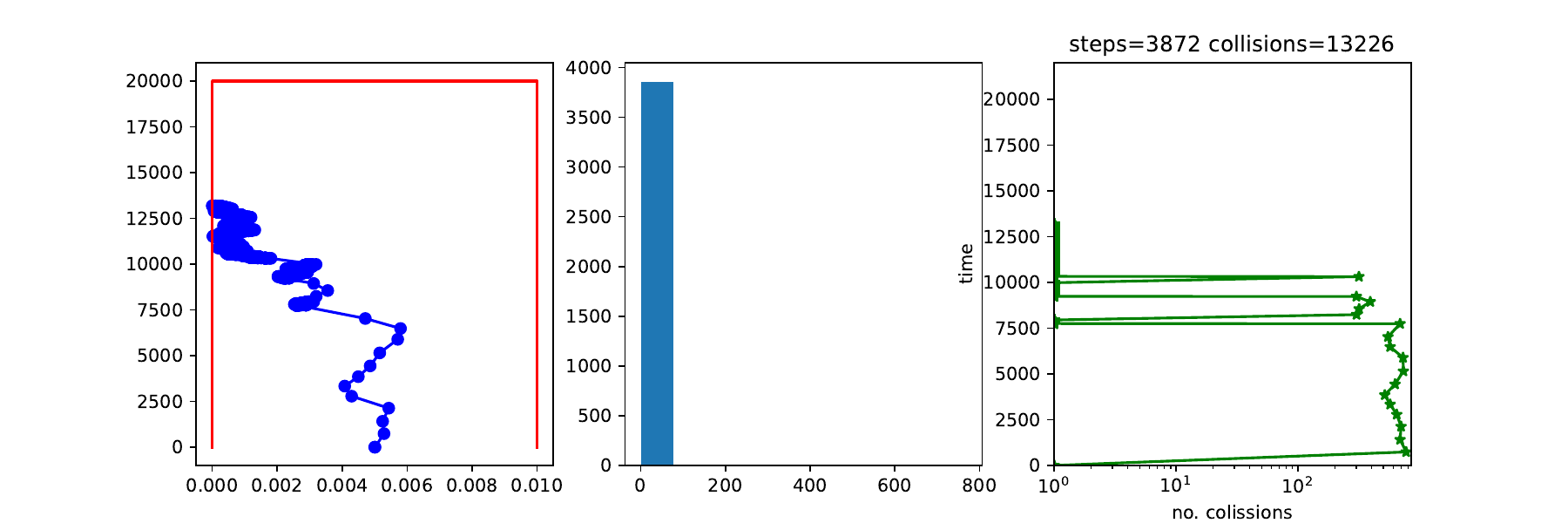}
	\caption{One example trajectory corresponding to results in figure	\ref{fig:1D_sn_laws_10_300}, i.e., $n_\Ncal=300$, $N=10$. Exit direction is $a^\star=-1$. Signification of the plots is as in figure \ref{fig:1D_sn_traj}.}
	\label{fig:1D_sn_traj_10_300}
\end{figure}

As a final test we will lower the  parameter $N$ to inquire whether the escape distributions are sensitive to the discretization of the number of directions in $\Dcal$. We take $N=10$, $n_\Ncal=300$, results are in figures~\ref{fig:1D_sn_laws_10_300}
and~\ref{fig:1D_sn_traj_10_300}. Of course, the distribution in the third column of figures~\ref{fig:1D_sn_laws_10_300} is very discrete (only $10$ directions are possible and some of them coincide). But the reassuring result is that the other two columns look very much like that in figure~\ref{fig:1D_sn_laws} which shows that the discrete nature of $\Dcal$ does not seem to play an important role in the shape of the escape distributions.

\section{Conclusions}

We presented a method that accelerates the sampling of the escape times, position and directions for particles undergoing collisions separated by exponentially long times.
The procedure works
by aggregating several collisions into a single step. The advantage of the method  is that it does not uses any model or Brownian walk approximation and therefore can treat in the same way a large range of collision parameters including those where the Brownian  walk approximation
(as used in \cite{fleck1984random,giorla1987random})
may not work well. The procedure operates by estimating conservatively the number of collisions that can safely be made before escaping the time-space domain; once this number estimated, the resulting position after those collisions is sampled exactly. The empirical results show that the method shows promising results for substantially diminishing the numerical cost while retaining excellent quality for the escape distribution.

The method was demonstrated in one dimension. 
Extensions to multi-dimensional domains or more general forms are left for future works. We have reasons to be optimistic because, in the worse case scenario, 
when the domain is a rectangle in high dimensions one can work by projecting on each of the dimensions and the setting will be that of a $1D$ segment; the number of aggregate collisions will be the minimum over each dimension; when the domain is not a rectangle one can consider the largest rectangle included in the domain. Finding even more efficient approaches in high dimensions is left for future work. 

\bibliographystyle{unsrt}

\end{document}